\documentclass[12pt]{article}

\renewcommand{\title}[1]{\null\vspace{25mm}
\noindent{\Large{\bf #1}}\vspace{10mm}
}
\newcommand{\authors}[1]{\noindent{\large #1}\vspace{20mm}
    }
\newcommand{\address}[1]{{\center{\noindent #1\vspace{10mm}}
    }}
\renewcommand{\abstract}[1]{\vspace{17mm}
\noindent{\small{\em Abstract.} #1}\vspace{2mm}
   }     
 
\newcommand{\nm}{\nonumber}
\newcommand{\us}{s}
\newcommand{\be}{\begin{equation}}
\newcommand{\ee}{\end{equation}}
\newcommand{\ba}{\begin{array}}
\newcommand{\ea}{\end{array}}
\newcommand{\bea}{\begin{eqnarray}}
\newcommand{\eea}{\end{eqnarray}}
\newcommand{\Li}{{\cal L}_\tau}
\newcommand{\id}{i_\tau}

\newcommand{\bzd}{\frac{2}{3}}
\newcommand{\bed}{\frac{1}{3}}
\newcommand{\tr}{{\rm tr}\,}
\newcommand{\del}{\delta_\tau}

\newcounter{saveeqn}

\begin{document}   \setcounter{table}{0}

\sloppy 

\begin{titlepage}
\begin{center}
\hspace*{\fill}{{\normalsize \begin{tabular}{r}
{\sf hep-th/0003222}\\
{\sf REF. TUW 00-10}\\
{\sf LYCEN 2000-30}\\
		      {\sf \today}\\
                              \end{tabular}   }}

\title{Interacting six-dimensional topological field theories}

\authors {F.~Gieres$^1$, H.~Nieder, T.~Pisar$^2$, L.~Popp 
and M.~Schweda$^3$}    \vspace{-20mm}
       
\address{Institut f\"ur Theoretische Physik,Technische Universit\"at Wien\\
      Wiedner Hauptstra\ss e 8-10, A-1040 Wien, Austria}
\footnotetext[1]{On sabbatical leave from Institut 
de Physique Nucl\'eaire de Lyon,  
Universit\'e Claude Bernard, 43, boulevard du 11 novembre 1918, 
F-69622-Villeurbanne.}
\footnotetext[2]{Work supported by the 
"Fonds zur F\"orderung der Wissenschaflichen Forschung", 
under Project Grant Number P11582-PHY.}
\footnotetext[3]{email: mschweda@tph.tuwien.ac.at}       
\end{center} 
\thispagestyle{empty}

\abstract{We study the gauge-fixing and symmetries 
(BRST-invariance and 
vector supersymmetry)  
of various six-dimensional 
topological models involving Abelian or non-Abelian 
$2$-form potentials.}

\end{titlepage}

\tableofcontents
\thispagestyle{empty}
\newpage
\setcounter{page}{1}

\section{Introduction}

Recently, L.~Baulieu and P.~West introduced 
 a six-dimensional topological model of Witten-type involving 
$2$-form potentials 
\cite{Baulieu:1998xd}. 
In the sequel, the gauge-fixing procedure and twist in this model
have been studied in more detail in reference
\cite{Ita:1999mx}; these authors also    
determined vector supersymmetry 
(VSUSY-) transformations which represent an additional 
symmetry of the model.

The goal of the present paper is to discuss two different 
generalizations of the free Abelian model \cite{Baulieu:1998xd}
to interacting models. 
The first consists of coupling the Abelian $2$-form potentials to 
a non-Abelian Yang-Mills field by virtue of a Chern-Simons term, 
as suggested in reference \cite{Baulieu:1998xd}. 
The corresponding action represents a six-dimensional 
topological version of the Chapline-Manton 
 term appearing in the action for ten-dimensional supergravity
 coupled to super Yang-Mills theory \cite{gsw}. 
The second generalization consists of considering   
non-Abelian (charged)
$2$-form potentials which are coupled to a Yang-Mills connection, 
following the lines of reference \cite{Baulieu:1998yx}.

Before discussing these generalizations, we summarize 
the free Abelian model \cite{Baulieu:1998xd,Ita:1999mx}
while using differential forms to simplify the notation
(section 2).
 Our paper concludes with some comments concerning 
 possible extensions of these six-dimensional 
 topological models. 
 We note that all of our considerations concern the 
classical theory (tree-level).

\section{Free Abelian model}

The arena is a compact pseudo-Riemannian $6$-manifold
${\cal M}_6$ 
and the basic fields are Abelian $2$-form potentials 
$B_2$ and $B^c_2$ which are independent of each other. 
From the associated curvature 
$3$-forms  
$G_3=dB_2$ and $G^c_3=dB^c _2$, 
one can construct the classical action  \cite{Baulieu:1998xd} 
\be
	\Sigma_{cl} = \int G_3 \, G_3^c . 
\label{freeinv}
\ee
Here and in the following, the integrals are understood as integrals
of $6$-forms over ${\cal M}_6$ 
and the wedge product symbol is always omitted. 

\subsection{Symmetries}

The action (\ref{freeinv}) is invariant under the 
{\em ordinary  gauge transformations}
\be
\label{2}
	\delta_{\lambda_1}B_2=d\lambda_1
\quad , \quad 
\delta_{\lambda_1}B_2^c= 0, 
\ee
which represent a reducible symmetry in the present case, 
and it is invariant under the {\em shift-} (or topological Q-) {\em
symmetry} \be
\label{3}
	\delta_{\lambda_2}B_2=\lambda_2
\quad , \quad 
\delta_{\lambda_2}B_2^c= 0,
\ee
which also represents a reducible symmetry. 
In equations (\ref{2}),(\ref{3}) and in the following, 
it is understood that the ``$c$-conjugated'' equations 
also hold, e.g. 
$c$-conjugation of equation  (\ref{2}) gives 
$\delta_{\lambda_1^c}B^c_2=d\lambda_1^c, \, 
\delta_{\lambda_1^c}B_2= 0$.

In the sequel, we will describe the infinitesimal symmetries 
in a BRST-framework and we will derive 
the BRST-transformations 
from a horizontality condition (Russian formula) 
\cite{Bertlmann}.
Thus, we introduce 
a series of ghost fields associated with the reducible gauge  
transformations (\ref{2}) 
and we collect them in a 
generalized $2$-form,
\be
	\tilde B_2=B_2+V_1^1+m^2.
\label{freeBexp}
\ee
Here, the upper and lower indices denote the 
ghost-number and form degree, respectively.  
The total degree of a field is the sum of its ghost-number  
and  form degree, and all commutators $[\cdot, \cdot ]$
are graded with respect to this total degree.  
The BRST-differential $s$, which describes both the ordinary 
gauge transformations and 
the shift-transformations, is combined with 
the exterior derivative $d$ in a single operator,
\bea
	\tilde d=d+\us,
\eea
which is nilpotent by virtue of the relations
$d^2 = s^2 =[s,d]=0$.
Thus, the generalized field strength 
\be
\tilde G_3 \equiv \tilde d\tilde B_2 
\label{HorFree}
\ee
satisfies the generalized Bianchi identity 
\be
	\tilde d\tilde G_3=0.
\label{dep}
\ee
The BRST-transformations of the 
classical and ghost 
fields\footnote{Here, ``classical'' fields
are not opposed to quantum fields, but simply  
refer to the fields appearing in the classical action.} 
are now obtained
from the {\em horizontality condition} \cite{Baulieu:1998xd,Ita:1999mx}
\be
	\tilde G_3=G_3+\psi_2^1+\varphi_1^2+\phi^3,
\label{freeFexp}
\ee
which involves a series of ghosts  associated with 
the  shift-symmetry. 
By inserting the field expansions (\ref{freeBexp}) 
and (\ref{freeFexp}) into relations (\ref{HorFree}) and 
(\ref{dep}), we obtain the $s$-variations \cite{Baulieu:1998xd,Ita:1999mx}
\be
\label{sbf}
\ba{rclcrcl}
s B_2&=&\psi_2^1-dV_1^1 
&,& 
s \psi_2^1 &=&-d \varphi_1^2
\nm \\
s V_1^1&=&\varphi_1^2-dm^2 
&,& 
s \varphi_1^2 &=&-d\phi^3 
\nm \\
s m^2&=&\phi^3 
&,&
s \phi^3&=&0
\ea
\ee
and $s G_3=-d \psi_2^1$. 
Since the field expansions (\ref{freeBexp}) and (\ref{freeFexp}) 
have not been truncated,
the obtained BRST-transformations are nilpotent by construction
\cite{first}.

\subsection{Gauge-fixing}

Let us briefly review the gauge-fixing procedure 
\cite{Baulieu:1998xd, Ita:1999mx}
while using differential forms.
To start with, we consider the shift 
degrees of freedom for the fields $B_2$ and $B_2^c$.
These are fixed by imposing a self-duality condition
relating the corresponding field strengths:
\be
\label{con}
* G_3 = - G_3^c .
\ee
This relation is equivalent to imposing a self-duality 
condition on $G_- \equiv d(B_2 - B_2^c)$ and an 
anti-self-duality condition on $G_+ \equiv d(B_2 + B_2^c)$.
Henceforth, relation (\ref{con}) is analogous to the 
self-duality condition for the curvature 
$2$-form $F$ in four-dimensional topological Yang-Mills theory
\cite{Witten:1988ze}.

With the help of a BRST-doublet 
$(\chi_3^{-1},H_3 )$, i.e. 
\be
	 \us \chi^{-1}_3= H_3 \quad , \quad \us H_3=0,
\label{BRSTantiH}
\ee
the constraint (\ref{con}) can be implemented in the 
gauge-fixing action:
\be
	\Sigma_{sd}
=  s \int  \{ \chi^{-1}_3
\left(*G_3+ G_3^c \right) \}.
\ee
Since the shift-symmetry represents a reducible symmetry,
it is necessary to re-iterate the gauge-fixing procedure
for the action $\Sigma_{cl}+\Sigma_{sd}$: this leads to the 
introduction of 
the anti-ghosts and Lagrange multipliers of tables \ref{SIXtabtopo} 
and \ref{SIXtabtopomult}, all of which have been  arranged in  
Batalin-Vilkovisky pyramids. These 
fields again represent BRST-doublets:
\be
\ba{rclcrcl}
	\us \phi^{-3}&=&\eta^{-2}&,& \us \eta^{-2}&=&0\\
	\us \varphi_1^{-2}&=&\eta_1^{-1}&,& \us \eta_1^{-1}&=&0\\
	\us \chi ^1&=&\eta^2&,&\us \eta^2&=&0.
\label{BRSTanti1}
\ea
\ee 
In summary, the  shift-invariance of the classical action 
is fixed by virtue of the gauge-fixing action  
\be
\Sigma_Q =  \Sigma_{sd} 
+ s \int  
\left\{ \varphi_1^{-2}d*\! \psi_2^1
+\phi^{-3}d*\! \varphi_1^2+\chi^1d*\! \varphi_1^{-2}
- CC\right\},
\label{gaugetopo}
\ee
where 
$CC$ stands for the $c$-conjugated expressions. 

The reducible gauge symmetry (\ref{2}) is fixed in a similar way:
one considers the usual gauge condition $d*\!B_2=0$
and re-iterates the gauge-fixing procedure. This leads to the 
introduction of the series of anti-ghosts and multipliers
presented in tables \ref{SIXtabgauge} 
and \ref{SIXtabgaugemult}, the 
$s$-variations being given by 
\be
\ba{rclcrcl}
	\us m^{-2}&=&\beta^{-1}&,&\us \beta^{-1}&=&0 \\
	\us V_1^{-1}&=&b_1&,&\us b_1&=&0 \\
	\us n&=&\beta^1&,&\us \beta^1&=&0.
\label{BRSTanti2}
\ea
\ee

\begin{table}[ht]
\setlength{\unitlength}{1cm}
\begin{center}
\begin{minipage}[t]{6cm}
\begin{center}
\begin{tabular}{ccccccc} 
        && $\psi^1_2$ \\
        & $\varphi^{-2}_1$&&$\varphi_1^2$ \\
        $\phi^{-3}$&&$ \chi ^1$&&$\phi^3$
\end{tabular}
\caption{Tower for shift ghosts  and anti-ghosts}
\label{SIXtabtopo}
\end{center}
\end{minipage}
\hfill
\begin{minipage}[t]{6cm}
\begin{center}
\begin{tabular}{ccccccc} 
        & $\eta^{-1}_1$& \\
        $\eta^{-2}$&&$\eta^2$&
\end{tabular}
\caption{Tower for shift multipliers}
\label{SIXtabtopomult}
\end{center}
\end{minipage}
\end{center}
\end{table}

\begin{table}[ht]
\setlength{\unitlength}{1cm}
\begin{center}
\begin{minipage}[t]{6cm}
\begin{center}
\begin{tabular}{ccccc} 
        && $B_2$ \\
        & $V^{-1}_1$&&$V_1^1$ \\
        $m^{-2}$&&$n$&&$m^2$
\end{tabular}
\caption{Tower for gauge ghosts and anti-ghosts}
\label{SIXtabgauge}
\end{center}
\end{minipage}
\hfill
\begin{minipage}[t]{6cm}
\begin{center}
\begin{tabular}{ccccccc} 
        & $b_1$& \\
        $\beta^{-1}$&&$\beta^1$&
\end{tabular}
\caption{Tower for gauge multipliers}
\label{SIXtabgaugemult}
\end{center}
\end{minipage}
\end{center}
\end{table}

\noindent 
Thus, the ordinary gauge degrees of freedom of $B_2$ 
are fixed by the functional
\be
	\Sigma_{og}= s \int  \{ V_1^{-1}d*\! B_2
+m^{-2}d*\! V_1^1+nd*\! V_1^{-1}-CC  \}
\label{gfB}
\ee
and the complete gauge-fixed action of the model reads 
\be
\Sigma
=\Sigma_{cl}+ \Sigma_Q+\Sigma_{og}.
\label{ActionFree}
\ee

\subsection{Vector supersymmetry}

Due to the fact that we are considering a topological model 
of Witten-type, one expects the complete gauge-fixed 
action to admit a VSUSY. At  the infinitesimal level, the 
VSUSY-transformations are 
described by the operator $\del$  
where $\tau \equiv \tau^\mu \partial_{\mu}$ 
is a constant, $s$-invariant vector field of ghost-number
zero\footnote{In order to avoid technical complications
related to the global geometry, we limit the considerations
of this section to flat space-time.}.
The variation $\del$ acts as an antiderivation which lowers the 
ghost-number by one unit and which anticommutes with $d$. 
The operators $s$ and $\del$  satisfy a   
graded algebra of Wess-Zumino type,
\be
	[\us,\del]=\Li ,
\label{algebra}
\ee
where $\Li \equiv [\id,d]$
denotes the Lie derivative along the vector field $\tau$ 
and $\id$ the
interior product by $\tau$. 
We will simply refer to the relation (\ref{algebra})
as the {\em SUSY-algebra}.

The $\del$-variations of  
all fields can be determined by applying 
the general procedure introduced in reference 
\cite{first}. To start with, we derive the VSUSY-transformations
of the classical and ghost fields
 by expanding the so-called 
{\em $0$-type symmetry conditions} 
\be
	\del \tilde B_2=0 
\quad , \quad 
 \del \tilde G_3=\Li \tilde B_2.
\label{sol1}
\ee
with respect to the ghost-number. We thus obtain 
\be
\ba{rclcrcl}
        \del B_2 & = & 0&, & \del \psi_2^1 &=& \Li B_2 \\
        \del V_1^1 &=& 0&, & \del \varphi^2_1 & = & \Li V^1_1 \\
        \del m^2 & = & 0&, & \del \phi^3 &=& \Li m^2.
\label{vectorsusy1}
\ea
\ee
The $\del$-variations of the anti-ghosts 
are found by requiring the $\del$-invariance
of the total action (\ref{ActionFree}) 
and by applying the commutation relations
(\ref{algebra}).
Finally, the  VSUSY-transformations of all multipliers
follow  from the ones of the corresponding anti-ghosts 
by imposing the algebra (\ref{algebra}) for all of them:
\be
\ba{rclcrcl}
	\del \chi^{-1}_3&=&0&, &\del H_3&=& \Li \chi_3^{-1} \\
	\del \phi^{-3}&=&0&, & \del \eta^{-2}&=& \Li\phi^{-3}\\ 
\del \varphi^{-2}_1&=&0&, & \del \eta_1^{-1}&=& \Li\varphi_1^{-2}\\ 	
	\del \chi^1&=&-\Li n
&, & \del \eta^2&=& \Li \chi^1+\Li\beta^1
\label{SUSYanti1}
\\
	\del m^{-2}&=&\Li\phi^{-3}&, 
& \del \beta^{-1}&=& \Li m^{-2}-\Li\eta^{-2}\\ 
	\del V_1^{-1}&=&\Li \varphi_1^{-2}&, 
& \del b_1&=& \Li V_1^{-1}-\Li \eta_1^{-1}\\ 
	\del n&=&0&, & \del \beta^1&=&\Li n. 
\ea
\ee
Thus, it is by construction that 
the total action is  
$\del$-invariant and that the SUSY-algebra is fulfilled
off-shell for all fields of the model. 
Our results coincide with those found in reference \cite{Ita:1999mx}
by other methods. We refer to the latter work for 
the relation of VSUSY to a twist of a supersymmetric field 
theory.

\section{Abelian model with Chern-Simons term}

The authors of reference \cite{Baulieu:1998xd}
considered the 
interaction of the Abelian $2$-forms $B_2$ and $B_2^c$ with 
a non-Abelian Yang-Mills (YM) connection $A$  by virtue of 
a Chern-Simons term 
with coupling constant $\lambda$, 
\be
\Omega_3 (A) = \lambda \, \tr (AdA+\bzd AAA).
\ee
The proposed action reads\footnote{We do not include
the ordinary YM-action $\int \tr (F *\!F)$ as in reference 
\cite{Baulieu:1998xd} since it depends on the metric
and therefore destroys the topological nature of the model.}
\be
\hat\Sigma_{cl} = \int \left(G_3 - \Omega_3 \right) 
\left(G_3^c - \Omega_3 \right) .
\label{int6d}
\ee
 This functional represents a six-dimensional topological version of the 
 expression $\int \tr (G_3 - \Omega_3) *\!(G_3 - \Omega_3)$
which appears in the action for ten-dimensional 
supergravity coupled to super YM \cite{gsw}. 

The equations of motion for $A$ and $B_2$ 
(or $B_2^c$) have the form 
\[
F (G_3 - G_3^c) =0 
\qquad {\rm and} \qquad \tr (FF) =0,  
\]
where $F=dA+{1\over 2}[A,A]$ denotes 
the curvature $2$-form associated to $A$.
The latter equations imply $F=0$ (i.e. the same equation of motion
as in the three-dimensional Chern-Simons theory). 

\subsection{Symmetries}

The action (\ref{int6d}) is not anymore invariant under the shift
$\delta B_2 = \psi_2^1$. However, it is invariant under the YM-gauge
transformations 
\bea
\delta A& = & -Dc \equiv-(dc+[A,c])
 \nm \\
\delta B_2 &=& \lambda \, \tr (cdA)
\, = \, \delta B_2^c ,
\eea
which leave $G_3 -\Omega_3$ and $G_3^c -\Omega_3$ invariant. 

The BRST-transformations of $A$ and of the YM-ghost 
read
\be
s A=-Dc 
\qquad , \qquad 
s c=-{1\over 2}[c,c].
\label{sa}
\ee
They  follow from the 
horizontality condition $\tilde F =F$, where
\be
\tilde F = \tilde d\tilde A+{1\over 2}[\tilde A,\tilde A]
\qquad , \qquad \tilde A=A+c . 
\label{CShorA}
\ee

The generalized Chern-Simons form 
\[
\tilde{\Omega} _3 = \lambda \, \tr ( \tilde A 
\tilde d \tilde A +
\bzd \tilde A \tilde A \tilde A)
\]
can be expanded with respect to the ghost-number, 
\be
	\tilde\Omega_3=\Omega_3+\Omega_2^1+\Omega_1^2+\Omega^3,
\ee
which provides the well-known solution of the descent equations
(e.g. see \cite{PiguetSorella}) 
\be
\begin{array}{rclcrcl}
\Omega_3&=&\lambda \, \tr (AdA+\bzd AAA ) 
\quad &,& \quad \us\Omega_3+d\Omega_2^1&=&0
 \\
\Omega_2^1&=&\lambda \, \tr (cdA ) 
\quad&,& \quad \us\Omega_2^1+d\Omega_1^2&=&0
\\
\Omega_1^2&=&\lambda \, \tr (-ccA ) 
\quad &,&\quad \us\Omega_1^2+d\Omega^3&=&0
 \\
\Omega^3&=&\lambda  \,\tr (-\bed ccc )
\quad &,& \quad	\us\Omega^3&=&0.
\end{array}
\label{descent}
\ee

We now use this result to discuss 
the $B$-field sector. The generalized field strength of $B_2$
is defined as before (i.e. 
$\tilde G_3 = \tilde d \tilde B_2$ with 
$\tilde B_2 = B_2 +V_1^1 + m^2$), but the horizontality condition
(\ref{freeFexp}) of the free model is now replaced by the 
{\em horizontality condition} 
\be
\tilde G_3 = G_3 + \Omega_2^1  
+ \Omega_1^2 + \Omega^3 .
\label{new}
\ee
Expansion with respect to the ghost-number yields the 
$s$-variations 
\bea
	\us B_2&=& -dV_1^1+\Omega_2^1 \nm \\
	\us V_1^1&=& -dm^2+\Omega_1^2 \nm \\
	\us m^2&=& \Omega^3, 
\label{sb}
\eea
where the explicit expressions for 
the $\Omega_p^q (A,c)$ were  
given in equations (\ref{descent}).
Furthermore, substitution of (\ref{new}) in 
$\tilde d \tilde G_3=0$ leads to 
$s G_3=- d\Omega_2^1$ (and reproduces some of the descent
equations (\ref{descent})).

The BRST-transformations  (\ref{sa}) and (\ref{sb})
leave the classical action (\ref{int6d}) invariant.

\subsection{Gauge-fixing}

In the YM-sector, the gauge symmetry is fixed in the standard
way,  
\be
\Sigma^A_{gf}=s \int  \tr \left\{ \bar c d*\! \! A \right\}
=\int  \tr\left\{bd*\! \! A
-\bar c d*\! \!Dc \right\},
\ee
where we made use of a BRST-doublet
$(s \bar c=b , \,  
s b=0)$.

In the $B$-sector, the local symmetry $\delta B_2 = -dV_1^1$ is 
fixed as for the free model, i.e. by introducing the gauge-fixing 
functional $\Sigma^B_{gf} \equiv \Sigma_{og}$
given by equation (\ref{gfB}).

In summary, the complete action of the interacting 
model reads 
$\Sigma_{int}=\hat\Sigma_{cl}+  \Sigma^B_{gf} +\Sigma^A_{gf}$.

\subsection{VSUSY}

Due to the absence of shift-symmetries in the present model, 
the only possible choice for VSUSY-transformations 
is given by the 
so-called {\em $\emptyset$-type symmetry conditions} \cite{first}.
However, the derivation of $\del$-variations for all 
fields is substantially more complicated in the present case
since the SUSY-algebra only closes on-shell. Therefore, we 
will not further elaborate on this point here \cite{pisar}.

\section{Non-Abelian model}

Consider a YM-connection $A$ 
and a $2$-form potential $B_2$, both with values in a given Lie 
algebra. The field strength of $B_2$ is now defined by 
\be
G_3 =DB_2 \equiv dB_2 + [A,B_2 ] 
\ee
and it satisfies the second Bianchi identity
$DG_3 = [F,B_2]$, 
where $F =dA + {1\over 2}[A,A]$ denotes the YM-curvature. 

A natural generalization 
of the  action (\ref{freeinv}) for the Abelian potentials is given by 
\cite{Baulieu:1998yx}
\bea
\Sigma_{cl}&=& \int  \tr \left\{G_3 \; G_3^c
-F \, [B_2, B_2^c ]\right\}.
\label{action}	
\eea
Neither $B_2$ nor $A$ propagate in this model 
(very much like $A$ in 
four-dimensional topological YM-theory).

\subsection{Symmetries}
   
Following reference \cite{Baulieu:1998yx}, 
we now spell out all local symmetries 
of the functional (\ref{action}).
As in the previously discussed models, one considers the 
generalized gauge fields 
\be
	\tilde A=A+c 
\qquad , \qquad 
	\tilde B_2=B_2+V_1^1+m^2
\ee
and the associated generalized field strengths 
\be
\tilde F = \tilde d \tilde A 
+ {1\over 2} [\tilde A, \tilde A ] 
\qquad , \qquad
\tilde G_3 =\tilde D \tilde B_2 
\equiv \tilde d \tilde B_2 +[\tilde A ,\tilde B_2 ] .
\ee 

The BRST-transformations in the YM-sector 
can be summarized by the following  
{\em horizontality condition} 
which involves ghost fields for 
the shifts of $A$:
\be
\tilde F=F+\psi_1^1+\varphi^2.
\label{h1}
\ee
From this relation 
and  the generalized Bianchi identity $\tilde D \tilde F =0 $, 
we obtain 
\be
\begin{array}{rclcrcl}
s A&=&\psi_1^1-Dc 
\quad &,& \quad 
s \psi_1^1&=&-D\varphi^2-[c,\psi_1^1]
\nm \\
s c&=&\varphi^2-{1 \over 2}[c,c] 
\quad &,& \quad 
s \varphi^2&=&-[c,\varphi^2]
\label{sbb}
\end{array}
\ee
and  $sF=-D\psi_1^1-[c,F]$.

In the $B$-sector, the $s$-variations follow from the 
{\em horizontality condition} \cite{Baulieu:1998yx}
\be 
\label{h2}
	\tilde G_3=G_3+\psi_2^1+\varphi_1^2+\phi^3
\ee
and  the generalized Bianchi identity 
$\tilde D \tilde G_3 = [\tilde F  ,\tilde B_2 ]$: they read 
\be
\begin{array}{rclcrcl}
	s B_2&=&\psi_2^1-DV_1^1-[c,B_2] 
\quad &,& \quad 
s\psi_2^1 &=&-D\varphi_1^2-[c,\psi_2^1]
+[F,m^2]+[\psi_1^1,V_1^1]+[\varphi^2,B_2] 
\nm \\
	s V_1^1&=&\varphi_1^2-Dm^2-[c,V_1^1] 
\quad &,& \quad 
s\varphi_1^2&=&-D\phi^3-[c,\varphi_1^2]
+[\psi_1^1,m^2]+[\varphi^2,V_1^1] 
\nm \\
	s m^2&=&\phi^3-[c,m^2]
\quad &,& \quad 
s\phi^3&=&-[c,\phi^3]+[\varphi^2,m^2]
\end{array}
\label{2BRSTB}
\ee
and  $s G_3=-D\psi_2^1-[c,G_3]+[F,V_1^1]+[\psi_1^1,B_2]$.
The action (\ref{action}) is inert under the BRST-transformations 
(\ref{sbb}),(\ref{2BRSTB}) which are nilpotent by construction.

\subsection{Gauge-fixing}
   
The shift- and  ordinary gauge symmetry 
in the $B$-sector are fixed as in the 
free Abelian model, except that all fields are now Lie algebra-valued. 
Thus, the total gauge-fixing action in the $B$-sector is given by
\be
	\Sigma^{B}= \Sigma^B_{Q} + \Sigma^B_{og}, 
\ee
with 
\bea
\Sigma_Q ^B& =& s \int \tr \left\{ \, \chi^{-1}_3
\left( *G_3 + G_3^c \right) 
\, + \, \left[ \varphi_1^{-2}d*\! \psi_2^1
+\phi^{-3}d*\!\varphi_1^2
+\chi^1d*\! \varphi_1^{-2}-CC \right] \right\}
\label{sdB} \\
\Sigma^B_{og}&=& s \int  \tr \left\{V_1^{-1}d*\! B_2
+m^{-2}d*\! V_1^1+nd*\! V_1^{-1}- CC\right\},
\nonumber
\eea
The BRST-transformations of the 
anti-ghost and multiplier fields are given by
equations  (\ref{BRSTantiH}),(\ref{BRSTanti1}) 
and (\ref{BRSTanti2}).

In the YM-sector, the gauge-fixing can be done 
along the lines of  
four-dimensional topological YM-theory.
However, the familiar  four-dimensional 
self-duality condition for the curvature form $F$
does not make sense in six dimensions and it  
has to be generalized by introducing a $4$-form $T_4$ 
which is 
invariant under some maximal subgroup of $SO(6)$ 
\cite{corrigan, bks}: the self-duality condition can then
be written as 
\be 
*F = \Omega_2 F 
\qquad {\rm with} \quad 
\Omega_2 \equiv *T_4.
\ee
This constraint is implemented by the gauge-fixing action 
\be
\Sigma^A_{sd} =  
s \int \tr \left\{ \chi_4^{-1} 
\left(* F - \Omega_2 F \right) \right\} , 
\ee
where $\chi_4^{-1}$ belongs to a BRST-doublet
$(s \chi_4^{-1}=H_4 , \, sH_4 =0)$. 
The residual gauge symmetries can then be fixed   
as in topological YM-theory by using a linear gauge-fixing 
term \cite{Brandhuber:1994uf, first}, 
\be
\Sigma ^A=  \Sigma^A_{sd}  
+ s \int \tr \left\{ \bar{\phi}^{-2}d*\!\psi_1^1
+\bar cd*\!A \right\},
\ee
which involves the BRST-doublets $(\bar{\phi}^{-2}, \eta^{-1})$ 
and $(\bar c , b)$. 
In summary, the total gauge-fixed action is given by
$\Sigma=\Sigma_{cl} + \Sigma^B + \Sigma^A $.

\subsection{VSUSY}
   
In order to derive the VSUSY-transformations,  
one considers {\em $0$-type symmetry conditions} in the 
$A$- and $B$-sectors:
\be
\begin{array}{rclcrcl}
	\del \tilde A &=&0
\quad &,& \quad 
\del \tilde F & =&\Li \tilde A
\\
\del\tilde B_2&=&0
\quad &,&\quad 
\del \tilde G_3& =& \Li\tilde B_2.
\end{array}
\label{vsAB}
\ee
By expanding with respect to the ghost-number and 
substituting the 
horizontality conditions (\ref{h1}),(\ref{h2}), one finds
\be
	\del (A, c)=0
\qquad , \qquad 
	\del (B_2,  V_1^1, m^2) =0.
\label{2susyA}
\ee
and 
\be
\begin{array}{rclcrcl}
\del \psi_1^1&=& \Li A \quad &,&\quad \del \psi_2^1 &=& \Li B_2
\\
\del \varphi^2&=&\Li c \quad &,&\quad \del \varphi^2_1 & = & \Li V^1_1
\\
& & & & \del \phi^3 &=& \Li m^2.
\end{array}
\ee

The $\del$-variations for the BRST-doublets 
occurring in the gauge-fixing action of the $B$-sector 
 are given by equations 
(\ref{SUSYanti1}) and those in the YM-sector read
\be
\ba{rclcrcl}	
\del \bar c&=&-\Li \bar{\phi} ^{-2}&, &\del b
&=& \Li \bar c+\Li \eta^{-1} \nm \\
\del \chi^{-1}_4&=&0&, & \del H_4&=& \Li \chi_4^{-1} \nm \\
\del \bar{\phi} ^{-2}&=&0&, & \del \eta^{-1}&=& \Li \bar{\phi} ^{-2}. 
\label{2susyAghost}
\ea
\ee
The total action is invariant under the given 
VSUSY-transformations which satisfy 
the VSUSY-algebra off-shell.

\section{Concluding remarks}

A possible generalization of the non-Abelian model consists 
of the addition of a $BF$-term $\int {\rm tr} \, (B_4F)$ 
(see \cite{PiguetSorella} and references therein for such models 
in arbitrary dimensions): such a term breaks the invariance 
under shifts of $A$. Other possible extensions \cite{Baulieu:1998yx}
are the 
inclusion of the topological invariant $\int \Omega_2 {\rm tr} \, (FF)$
(where $\Omega_2$ is a closed $2$-form), which leads to ``nearly 
topological'' field theories \cite{bks},
or the addition of a term $\int {\rm tr} \, (F\, DZ_3)$
involving a $3$-form $Z_3$.
The resulting models can be discussed along the lines of the present 
paper \cite{pisar}.


\bigskip 
\bigskip

\providecommand{\href}[2]{#2}\begingroup\raggedright\endgroup

\end{document}